\documentstyle[11pt,epsfig]{article}
\begin{document}
\renewcommand{\thefootnote}{\fnsymbol{footnote}}
\sloppy
\newcommand{\rp}{\right)}
\newcommand{\lp}{\left(}
\newcommand \be  {\begin{equation}}
\newcommand \bea {\begin{eqnarray}}
\newcommand \ee  {\end{equation}}
\newcommand \eea {\end{eqnarray}}

\title{Download relaxation dynamics on the WWW following newspaper
publication of URL}
\thispagestyle{empty}

\author{Anders Johansen$^1$ and Didier Sornette$^{1,2,3}$ \\
$^1$ Institute of Geophysics and
Planetary Physics\\ University of California, Los Angeles, California 90095\\
$^2$ Department of Earth and Space Science\\
University of California, Los Angeles, California 90095\\
$^3$ Laboratoire de Physique de la Mati\`{e}re Condens\'{e}e\\ CNRS UMR6622 and
Universit\'{e} de Nice-Sophia Antipolis\\ B.P. 71, Parc
Valrose, 06108 Nice Cedex 2, France}

\date{}

\maketitle

\begin{abstract}

A few key properties of the World-Wide-Web (WWW) has been established
indicating the lack of any characteristic scales for the WWW, both in its
topology and in its dynamics. Here, we report an experiment which
quantifies another power law describing the dynamical response of the WWW
to a Dirac-like perturbation, specifically  how the popularity of a web
site evolves and relaxes as a function of time, in response to the
publication of a notice/advertisement in a newspaper. Following the
publication of an interview of the authors by a journalist which contained
our URL, we monitored the rate of downloads of our papers and found it to
obey a $1/t^b$ power law with exponent $b=0.58\pm 0.03$. This small exponent
implies long-term memory and can be rationalized using the concept of
persistence, which specifies how long a relaxing dynamical system remains
in a neighborhood of its initial configuration.

\end{abstract}

\pagenumbering{arabic}

\newpage

It is generally accepted that the World-Wide-Web (WWW) provides one of
the most efficient methods for retrieving information. However, little is known
about how information actually flows through the WWW and even less on how
the WWW interacts with other types of media. Most studies have until now
focused on statistical properties of the WWW and the people surfing on it,
the ``internauts''.
A few key properties has been established indicating the lack of any
characteristic scales for the WWW \cite{IMA}: (i) the distribution of the
number of pages per site is an approximate power law \cite{Hub1}; (ii)
the distributions of outgoing (Uniform Resource Locator or
URLs found on an HTML document) and incoming
(URLs pointing to a certain HTML document) links are well-described by
a universal power law which seems independent of the search engine
\cite{diameter};
(iii) the distribution of independent hits or users per web-site also seems
to follow a power law and the ranking of sites according to their popularity
is well-described by Zipf's law \cite{Hub2,Hub3}; (iv) the distribution of
waiting times to access a given page is also a power law distribution
\cite{willinger1,willinger2} and the correlation function of the WWW traffic
intensity as a function of time also exhibits a slow power law decay
\cite{willi3}.

These properties are believed to reflect the evolutionary self-organizing
dynamics
of the WWW, which is not well-understood and the subject of active
research \cite{IMA}.
The WWW provides in particular a very interesting proxy of a fast evolving
ecology
of heterogeneous agents in which several different times scales appear ranging
from  the largest time scale
corresponding to a significant evolution of the web network (months to years),
the response adjustment time of agents to network evolution
or to novel information (hours to months)
to the access times (seconds to minutes) of single WWW pages.

Here, we report an experiment which probes a property belonging to the
intermediate
time scale. Specifically, we quantify how the popularity of a web site evolves
and relaxes as a function of time, in response to the publication of a
notice/advertisement in
a newspaper. The authors were interviewed by a journalist
from the Danish newspaper JyllandsPosten on a subject of rather broad and
catchy interest,
namely stock market crashes. The interview was
published on the 14 April 1999 in both the paper version of the newspaper
as well as in the electronic version (with access restricted to subscribers)
and included
the URLs where the authors' research papers on the subject could be retrieved.
Specifically, the URLs were the search engine of the Los Alamos preprint
server
and the URL of the first author's home-page at the Niels Bohr Institute's
web-site.
Naturally, we had no means of monitoring the downloads from the Los Alamos
preprint server. However, all WWW-activity on the Niels Bohr Institute's
web-site
is continuously logged and kept for record. It was hence possible to monitor
the number of downloads of papers as a function of time.

Since the interview was published in Danish, the experiment
only probes a small fraction of the internauts, namely those capable
of reading Danish, thus essentially people of Danish, Icelandic,
Norvegian and Swedish origin and their immediate surroundings.
The results reported below have not been reproduced as the ``impact'' by
the publication of the interview provides a rather unique opportunity to
monitor
in real time the dynamics of information spreading and persistence. The
statistical significance can thus be improved in principle by repeating
this experiment
several times.

\begin{figure}
\begin{center}
\epsfig{file=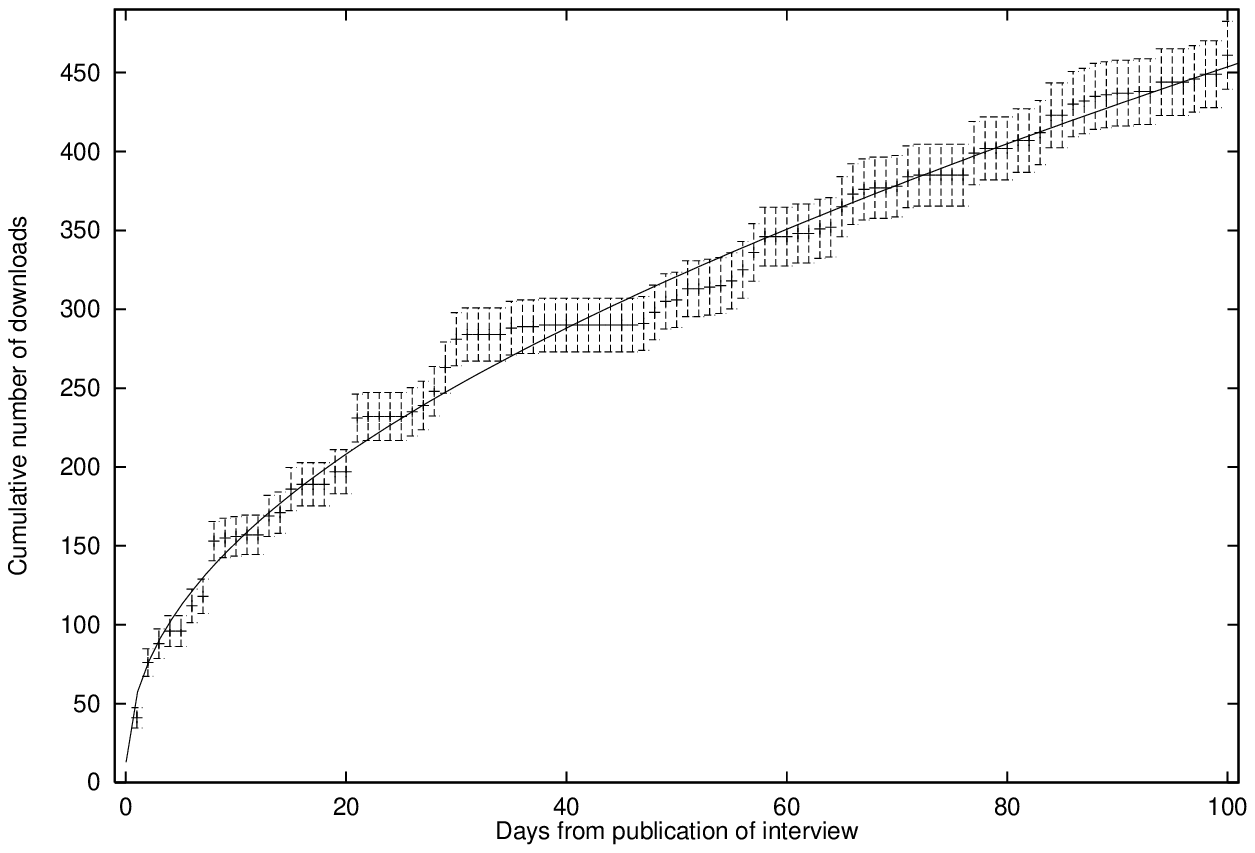,height=6.1cm}
\caption{\protect\label{hitsfit} Cumulative number of downloads $N$ as a
function of time $t$ from the appearance of the interview on Wednesday
the 14 April 1999. The fit is $N(t) = \frac{a}{1-b} t^{1-b} + ct$ with $b
\approx 0.58$
.}

\vspace{5mm}

\epsfig{file=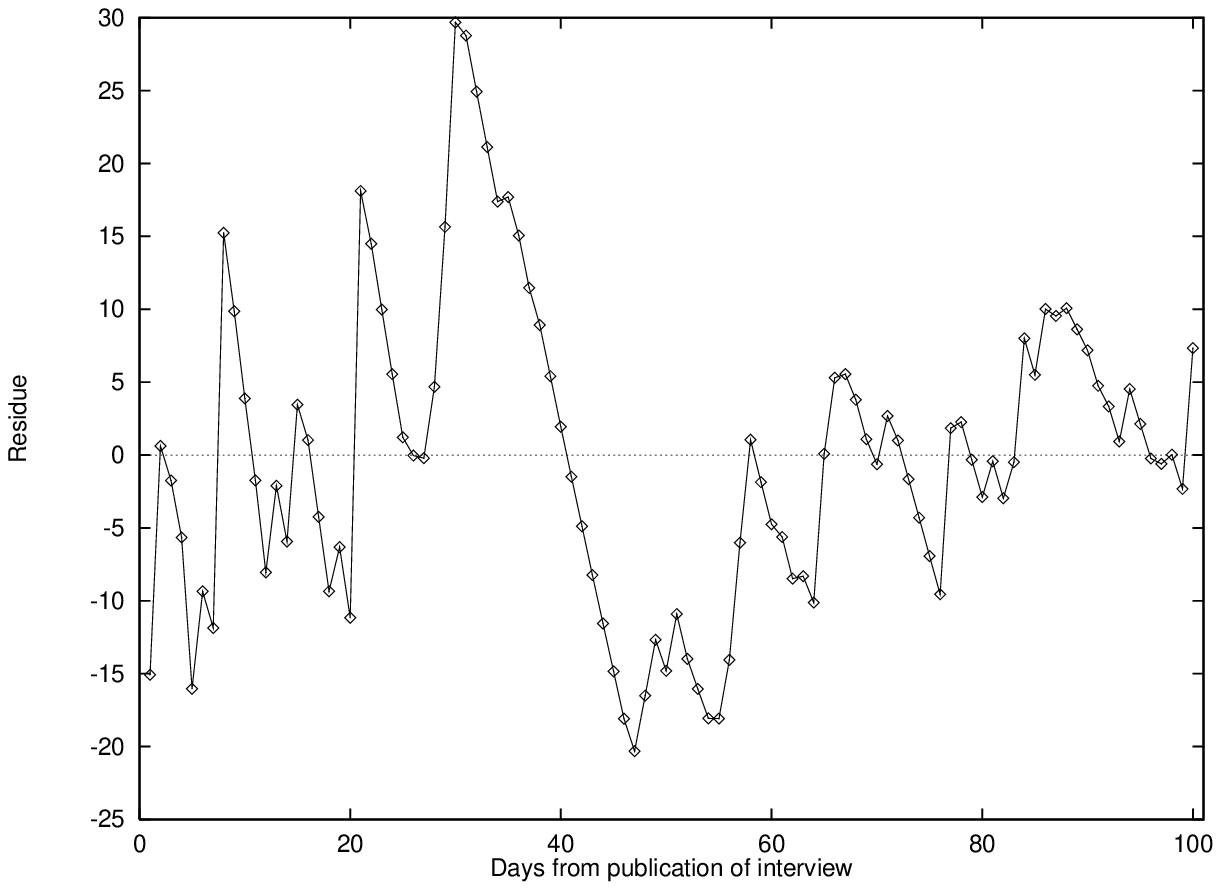,height=6.1cm}
\caption{\protect\label{hitsfitres} Residue obtained by subtracting the
fit shown in figure \ref{hitsfit} from the data points.}

\vspace{5mm}

\epsfig{file=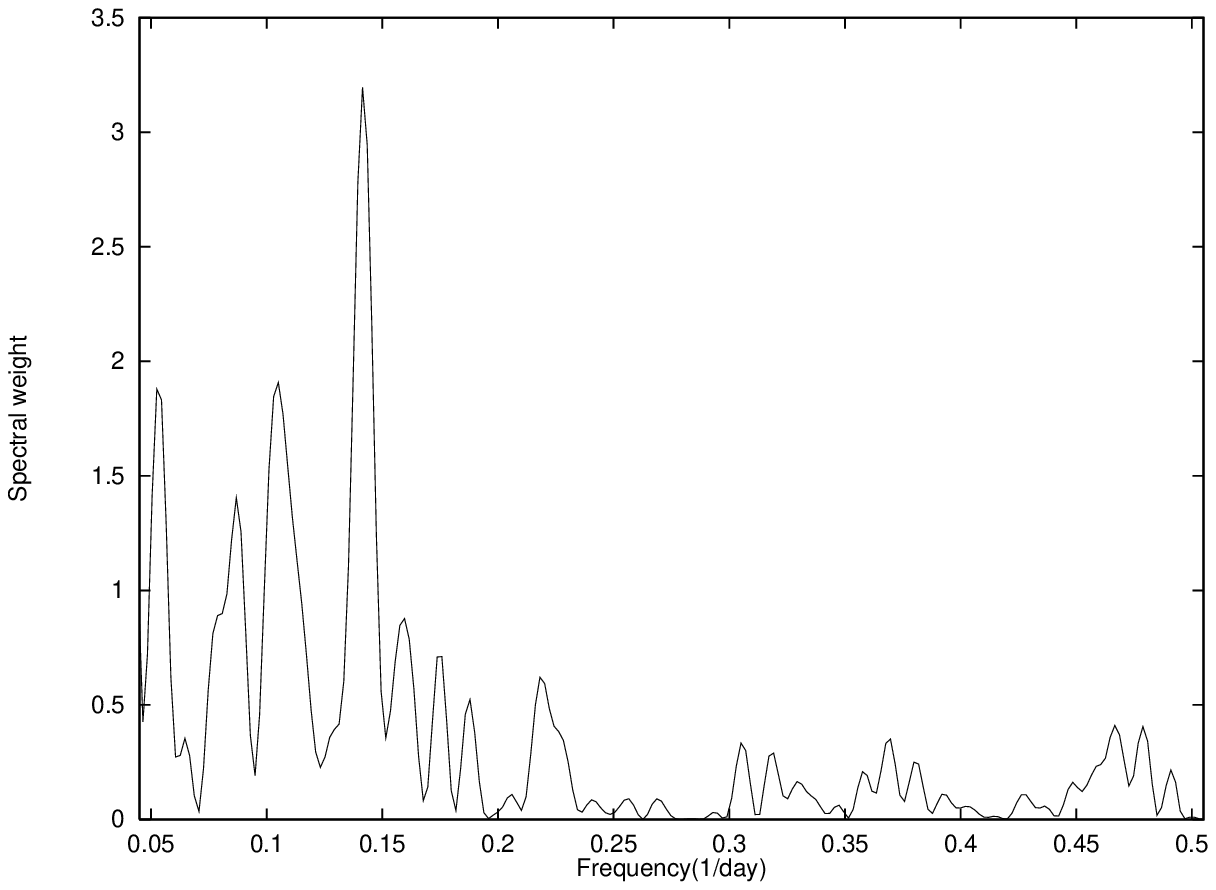,height=6.1cm}
\caption{\protect\label{hitsfitfp} Spectrum of data in figure \ref{hitsfitres}
showing a weekly periodicity. }
\end{center}
\end{figure}

In figure (\ref{hitsfit}), we show the cumulative number of downloads $N$
as a
function of time $t$ since the publication of the interview. Only downloads of
papers already posted on the home-page at the time of the publication of the
interview has been included in the count in order to keep the experiment as
well-defined as possible. The error-bars are taken as the square-root of
the number. We see that the data is surprisingly well-captured over
two decades by the relation
\be \label{downeq}
N\lp t\rp = \frac{a}{1-b} t^{1-b} + ct ,
\ee
corresponding to a download rate $dN(t)/dt = a t^{-b} +c$ giving the number
of downloads per unit time at a time $t$ after the publication of the
interview. The constant background rate $c$ takes into account downloads
from people unaware of the interview as well as robots.
The best fit parameters are $a= 23.1 \pm 0.5$ days$^{-1}$, $b\approx 0.58 \pm
0.03$ and $c \approx 0.76 \pm 0.31$ days$^{-1}$, over a total time
interval of $100$ days. Expression (\ref{downeq}) thus establishes
a novel self-similar relationship for the dynamical behavior on the WWW,
describing the slow relaxation of the system after an essentially
Dirac-like excitation. The coefficient $a$ controls the absolute number of
downloads per unit time and is thus not universal. It reflects the size of the
internaut population which is concerned by the experiment. Similarly, the
coefficient $c$ controls the background rate and depends on 1) how easily the
page can be found and 2) the general interest of the subjects posted on the
page.

The finding that the relaxation exponent $b$ is less than one has an important
consequence, namely non-stationarity and ``aging'' in the technical sense of
a breaking of ergodicity.
Consider $N$ successive downloads separated in time by $\Delta t_i,
i=1,...,N ,$
where
$\Delta t_1+\Delta t_2+ ...+\Delta t_N = t = N \langle \Delta t \rangle$. The
distribution of downloads time intervals $\Delta t$ is a power law
$1/\Delta t^{1+x}$, where $x$ is determined from the fact that
$$
\langle \Delta t \rangle \sim \int_0^{\Delta
t_{max}} d\tau {\tau \over \tau^{1+x}} \sim \Delta t_{max}^{1-x}~.
$$
 Since the maximum $\Delta t_{max}$ among $N$ trials is typically given by
$N \int_{\Delta t_{max}}^{\infty} {d\tau' \over \tau'^{1+x}} \sim 1$, we
have $\Delta
t_{max} \sim N^{1 \over x}$. Thus $t = N \langle \Delta t \rangle \sim N^{1
\over x}$ giving $N \sim t^x$,  for $x<1$.  We can thus identify the exponent
$x$ with $1-b$ and thus find that the distribution of waiting times between
successive downloads is a power law with an exponent $b \approx 0.58$ less
than one.
One can then show that this power law distribution of time intervals between
downloads implies that the longer
since the last download, the longer the expected time till the
next one \cite{Davis}. In other words, any expectation of a download that is
estimated today depends on the past in a manner which does not decay. This
is a hallmark of ``aging''. The mechanism is similar to the ``weak
breaking of ergodicity'' in spin glasses that occurs when the
exponent $x$ of the distribution of trapping times in meta-stable
states is less than one \cite{Bouchaud}.

How can we rationalize this relation (\ref{downeq})? We propose the
following very naive but illustrative model:
think of the population of internauts as subjected to the influence of the
newspaper publication
that may trigger an activity (downloading from our site). Let us think of this
influence as a field that diffuse and spread dynamically in the complex
space network of internauts and in their mind. This diffusive field
captures the dynamics of information, rumor spreading, psychological
decision and so on. Let us assume that the decision to act and download
from our
site is triggered when the influence field reaches a threshold. Then the
rate $dN(t)/dt$ is proportional to the probability for the field not to have
reached
the threshold, i.e. to the probability to remain in the neighborhood of its
initial state. This problem falls in the class of
the so-called ``persistence phenomenon''
discovered in a large variety of
systems \cite{persistence,Majumdar}, and
which specifies how long a relaxing dynamical system remains
in a neighborhood of its initial configuration. For a Gaussian process,
the persistence exponent $x$ can be shown to be a functional of the two-point
temporal correlator \cite{persistence,Majumdar,Sire}. For Markovian or weakly
non-Markovian random walk processes, the exponent $x$ and therefore $b$ is
close
to $1/2$, as we find empirically.

Figure (\ref{hitsfitres}) shows the residue obtained by subtracting the
formula (\ref{downeq}) from the data points. Figure (\ref{hitsfitfp})
shows that the spectrum of this residue is sharply peaked on a
characteristic frequency corresponding exactly to a period of one week.
Since the publication date of 14 April was a Wednesday, the dips shown in
figure (\ref{hitsfitres}) corresponds to Weekends: Apparently, most people
probed by the experiments still mainly have Internet and printer access
through their job, which explains the low activity during Weekends and
the weekly periodicity.

\newpage

\end{document}